\documentclass[dvips,aoas,preprint]{imsart}

\usepackage{amsthm,amsmath,latexsym, txfonts}
\usepackage{bm}
\usepackage{graphicx}
\usepackage{color}
\usepackage{rotating}

\usepackage[normalem]{ulem}
\usepackage{comment}
\usepackage{subfigure}

\RequirePackage[OT1]{fontenc}
\RequirePackage{amsthm,amsmath}
\RequirePackage{natbib}
\RequirePackage[colorlinks,citecolor=blue,urlcolor=blue]{hyperref}

\startlocaldefs

\theoremstyle{plain} \newtheorem{ass}{Assumption}

\newcommand{\bX}{\mathbf{X}}
\newcommand{\bY}{\mathbf{Y}}

\newcommand{\bZ}{\mathbf{Z}}
\newcommand{\bW}{\mathbf{W}}
\newcommand{\bA}{\mathbf{A}}
\newcommand{\bS}{\mathbf{S}}

\newcommand{\Aobs}{A_i^{obs}}
\newcommand{\Amis}{A_i^{mis}}
\newcommand{\Wobs}{W_i^{obs}}

\newcommand{\Yobs}{Y_i^{obs}}
\newcommand{\Ymis}{Y_i^{mis}}

\newcommand{\Uset}{{\cal{U}}_{s_0}}

\newcommand{\balpha}{\bm{\alpha}}

\newcommand{\bbeta}{\bm{\beta}}
\newcommand{\bgamma}{\bm{\gamma}}
\newcommand{\bmu}{\bm{\mu}}
\newcommand{\btheta}{\bm{\theta}}

\newcommand{\one}{\mathbf{1}}

\DeclareMathOperator\bE{\mathbb E} 

\endlocaldefs

\begin{document}

\begin{frontmatter}

\title{Evaluating the Causal Effect of University Grants on Student Dropout: Evidence from a Regression Discontinuity Design Using Principal Stratification}
\runtitle{Evaluating the Causal Effect of University Grants on Student Dropout}


\begin{aug}
\author{\fnms{} \snm{Fan Li}\thanksref{t3,m2}\ead[label=e3]{fli@stat.duke.edu}},
\author{\fnms{} \snm{Alessandra Mattei}\thanksref{t2,m1}\ead[label=e1]{mattei@disia.unifi.it}}
\and
\author{\fnms{} \snm{Fabrizia Mealli}\thanksref{m1}
\ead[label=e2]{mealli@disia.unifi.it}
}

\thankstext{t2}{Part of this paper was written when Alessandra Mattei was a senior fellow of the Uncertainty Quantification program of the U.S. Statistical and Applied Mathematical Sciences Institute (SAMSI).}
\thankstext{t3}{Supported by U.S. NSF grants SES-1155697 and SES-1424688}
\runauthor{Li, Mattei, Mealli}

\affiliation{Duke University\thanksmark{m2} and University of Florence\thanksmark{m1}}

\address{Alessandra Mattei and Fabrizia Mealli\\
University of Florence\\
Department of Statistics, Informatics, Applications\\
Viale Morgagni, 59 \\
50134 Florence, ITALY\\
\printead{e1}\\
\phantom{E-mail:\ }\printead*{e2}}

\address{Fan Li\\
Duke University\\
Department of Statistical Science\\
122 Old Chemistry Building		\\
Durham,  North Carolina 27708-0251, U.S.A.\\
\printead{e3} 
}
\end{aug}

\begin{abstract}
Regression discontinuity (RD) designs are often interpreted as local randomized experiments: a RD design can be considered as a randomized experiment for units with a realized value of a so-called forcing variable falling around a pre-fixed threshold. Motivated by the evaluation of Italian university grants, we consider a fuzzy RD design where the receipt of the treatment is based on both eligibility criteria and a voluntary application status. Resting on the fact that grant application and grant receipt statuses are post-assignment (post-eligibility) intermediate variables, we use the principal stratification framework to define causal estimands within the Rubin Causal Model. We  propose a probabilistic formulation of the assignment mechanism underlying RD designs, by re-formulating the Stable Unit Treatment Value Assumption (SUTVA) and making an explicit local overlap assumption for a subpopulation around the threshold. A local randomization assumption is invoked instead of more standard continuity assumptions. We also develop a model-based Bayesian approach to select the target subpopulation(s) with adjustment for multiple comparisons, and to draw inference for the target causal estimands in this framework. Applying the method to the data from two Italian universities, we find evidence that university grants are effective in preventing students from low-income families from dropping out of higher education.
\end{abstract}


\begin{keyword}
\kwd{Bayesian}
\kwd{Causal effects}
\kwd{Intermediate variables}
\kwd{Principal stratification}
\kwd{Randomization}
\kwd{Regression discontinuity}
\kwd{University grants}
\end{keyword}

\end{frontmatter}

\section{Introduction}
Amid the recent economic crisis in Europe, there has been a heated debate on how to arrange college students financial support, especially in terms of the instruments used, e.g., loans, grants, tuition waiver. Accurate evaluation of the effectiveness of the existing financial aid systems is crucial for providing information to policy makers to choose between different instruments. In Italy state universities offer financial aid every year to a limited number of eligible freshmen. The main objective of this intervention is to give equal opportunity to achieve higher education to motivated students irrespective of their economic background. Dropout from university is a relevant phenomenon in Italy: indeed, the low rate of university graduates among Italian youths is mainly due to the high dropout rate (about 30\%) rather than to a low enrollment rate. In this paper, we will investigate the causal effects of Italian university grants on preventing students from low-income families from dropping out of higher education, using data on first-year enrollees from two state universities.

In the Italian university system, only students who both meet a pre-fixed eligibility criteria and apply for a grant can receive the grant, consisting of tuition waiver, free meals and accommodation, and a limited amount of money around $3\,000$ Euros. The eligibility status depends on an economic measurement of the student's family income and assets falling below or above a pre-determined threshold. This allocation rule motivates us to adopt the regression discontinuity (RD) design framework for evaluation. RD design---a quasi-experimental design for causal inference---was first introduced in psychology by \cite{Thistle60} and has became increasingly popular since the late 1990s in economics and other fields. Recent surveys can be found in \cite{Cook08,ImbensLemieux08, Klaauw08,LeeLemieux10}. There are two general setups in RD designs, the sharp and the fuzzy RD designs. In the sharp RD design, the original form of the design, the treatment status is assumed to be a deterministic step function of a so-called \textit{forcing variable} or \textit{running variable}. All units with a realized value of the forcing variable on one side of a pre-fixed threshold are assigned to one regime and all units on the other side are assigned to the other regime. The basic idea underlying a RD analysis is that one can compare units with very similar values for the forcing variable, but different levels of treatment, to draw inference on the causal effect of the treatment at the threshold. Examples of sharp RD designs can be found, among others, in 
\cite{BerkLeuuw99, Lee08, MealliRampichini12}.
In the fuzzy RD design, the realized value of the forcing variable does not alone determine the receipt of the treatment, although a value of the forcing variable falling above or below the threshold acts as an encouragement or incentive to participate in the treatment. In those cases, the receipt of the treatment depends also on individual choices, which may confound treatment receipt. \cite{HTV01} establish a connection between fuzzy RD designs and the instrumental variable (IV) settings, and show that in a fuzzy RD  setting one can identify the local average treatment effect \citep{Imbens94} for a subpopulation of compliers at the threshold. Examples of fuzzy RD designs can be found, among others, in \cite{Klaauw02, BattistinRettore08, Garibaldi12}.

The Italian university grant allocation rule defines a fuzzy RD design because not all eligible students get a grant, they must apply first, and application is voluntary. Also ineligible students may apply, even if they will not receive any grant. Comparing to standard fuzzy RD designs where only assignment (eligibility) and receipt of the treatment (grant) are available, the additional data on the application status in this study can provide valuable information with important policy implications. In this article, we will show how to capitalize on application. In particular, a main methodological contribution of this article is to develop a framework for RD analysis that is fully consistent with the Rubin Causal Model \cite[RCM, ][]{Rubin74, Rubin78} using potential outcomes. Resting on the fact that grant application and grant receipt statuses are post-assignment (post-eligibility) intermediate variables, we adopt the principal stratification framework \citep{Frangakis02}---a generalization of the IV approach to noncompliance \citep{Angrist96,Imbens97}---to define causal estimands and lay the basis for inference.

In the literature, causal inference in RD designs is usually based on comparisons of units with close but distinct values of the forcing variable and relies on smoothness assumptions about the relationship between outcomes and the forcing variable around the threshold, which imply randomization at the single threshold value. For example, in fuzzy RDs, estimands are usually specified as ratio of differences of regression functions at the threshold, and inference generally relies on asymptotic approximations \citep[e.g.][]{ImbensLemieux08}.
In real applications, large-sample approximations might be unreliable due to the small sample size, and exact inference would be preferable. RD designs have been often described as designs that lead to locally randomized experiments around the threshold \citep{Lee08, LeeLemieux10, DinardoLee11}. Expanding on this interpretation, a recent strand of the literature \citep[e.g.,][]{Cattaneo15, Sales13} is moving towards a formal and well-structured definition of the conditions under which RD designs are equivalent to local randomized experiments.

We further develop the idea of local randomization; our goal is to provide a formal definition of the hypothetical experiment underlying RD designs, based on a description of the assignment mechanism, i.e., the process that describes why some units got assigned to different treatments, formalized as a unit-exchangeable stochastic function of covariates and potential outcomes.  The core of our framework is to assume there exists at least one subpopulation around the threshold where a \textit{local overlap} assumption holds.  For this subpopulation we explicitly introduce a \textit{local randomization} assumption.

Though our framework is not tied to any mode of inference, we choose the Bayesian approach for reasons explained later. In particular, a second methodological contribution of this article lies in our development of a Bayesian hierarchical modeling approach to adjust for multiple comparisons in selecting the target subpopulation(s). Our work contributes to the limited literature on Bayesian analysis to RD \citep{ChibJacobi11, ChibGreenberg14}, as well as to the literature on Bayesian causal inference \citep[e.g.,][]{Rubin78, Imbens97, Barnard03,Elliott10,Schwartz11,MLM13}.

In Section \ref{sec:estimand}, we introduce the basic setup and the causal estimands. In Section \ref{sec:RDD}, we propose a probabilistic formulation of the assignment mechanism for general RD designs, explicitly formulating the key assumptions, and elaborate it for the particular RD design used in the Italian university grants. Selection of the subpopulations where these assumptions hold is also discussed. A Bayesian approach for inferring causal effects in RD designs is developed in Section \ref{sec:Bayesian}. We then apply the proposed approach to evaluate causal effects of Italian university grants on student dropout in Section \ref{sec:Application}. Section \ref{sec:Conclusion} concludes.

\section{Causal estimands} \label{sec:estimand}
\subsection{Basic setup} \label{sec:SUTVA}
We introduce the notation in the context of Italian university grants.
Let $Z$ be the eligibility status, which is the initial assignment and plays the role of an ``instrument'' or an ``encouragement'' as in randomized experiments with noncompliance. Consider a sample or population of $N$ units; each can be either eligible to receive a treatment, $z =1$, or ineligible, $z=0$. In the Italian grants system, eligibility depends on the value of a combined measurement of one's assets including income and properties, adjusted for family size, denoted by $S$. If a student, satisfying preliminary grade criteria, has a value of $S$ falling below a pre-determined threshold, e.g. $s_0=15 \, 000$ euro, he/she is eligible, and not otherwise. That is, the eligibility status $Z_i$ for student $i$ is a deterministic function of $S$: $Z_i=\one(S_i \leq s_0)$, where $\one(\cdot)$ is the indicator function. Using the terminology in RD designs, $S$ is the forcing variable.

All variables measured after each unit $i$ is assigned eligibility $Z_i$, namely, the application status, the receipt of the grant and the dropout status, are post-assignment variables, and, in principle, eligibility may affect them. Thus we can define potential outcomes for these variables: for each student $i$ ($i=1,\ldots, N$), given eligibility status $z$ ($z=0,1)$, let $A_i(z)$ be an indicator for the potential grant application status (equal to $1$ if student $i$ applies for a grant and $0$ otherwise), $W_i(z)$ be an indicator for the potential treatment received (equal to $1$ if student $i$ receives a grant and $0$ otherwise), and $Y_i(z)$ be the potential indicator for dropout ($1$ if student $i$ drops out of university, $0$ otherwise). These notations, with only two potential outcomes for each post-treatment variable for each unit, reflect the acceptance of the Stable Unit Treatment Value Assumption \cite[SUTVA,][]{Rubin80}, which implies that there is no interference between units and that there are no levels of the eligibility status other than zero and one. A more explicit formulation of SUTVA will be introduced in Section \ref{sec:AM}.

For each unit, $i$, given the observed eligibility status $Z_i$, the following variables are observed:  $A^{obs}_i = A_i(Z_i)$, the observed application status; $W^{obs}_i = W_i(Z_i)$, the observed treatment received; and $Y^{obs}_i=Y_i(Z_i)$, the observed dropout status. The remaining potential outcomes are missing: $A^{mis}_i = A_i(1-Z_i)$, $W^{mis}_i = W_i(1-Z_i)$, and $Y^{mis}_i=Y_i(1-Z_i)$. A vector of $p$ pre-treatment variables, $\bX_i$, is also observed for each unit. We use boldface upper-case letters to denote the vector of values of a variable for all units from hereon. For example, $\bZ =(Z_1, \ldots, Z_N)'$, $\bA^{obs}=(A_1^{obs},\ldots,A_N^{obs})'$.

\subsection{The role of Principal Stratification for causal inference in Fuzzy RD designs} \label{sec:PS}

In the RCM, a causal effect is defined as a comparison of the potential outcomes $Y_i(1)$ and $Y_i(0)$, e.g., $\bE[Y_i(1)-Y_i(0)]$, for \emph{a common set of units}. Obviously, in our study, such comparisons between potential dropout statuses only measure the effect of the eligibility status. To draw inference about the causal effect of \emph{receiving a grant}, additional structure and assumptions are required. Since both the application status and receipt of the grant are post-assignment intermediate variables,  we adopt the Principal Stratification \citep{Frangakis02} framework.

For each intermediate variable, principal stratification defines a cross-classification of subjects into groups, named principal strata, defined by the joint potential values of that intermediate variable under each of the assignments being compared. In our study, based on the application status $A$, subjects are classified into four (latent) principal strata, $G_i \equiv (A_i(0), A_i(1))$: compliant-applicants $G_i=(0,1)=CA$, students who would not apply if ineligible, but would apply if eligible; always-applicants $G_i=(1,1)=AA$, students who would apply irrespective of their eligibility status; never-applicants $G_i=(0,0)=NA$, students who would not apply  irrespective of their eligibility status; and defiant-applicants $G_i=(1,0)=DA$, students who would not apply if eligible, but would apply if ineligible. Because principal strata are not affected by assignment, we can define \textit{population-average} causal effects conditional on the principal strata, known as principal causal effects:
\begin{eqnarray}
\tau_g \equiv \bE[Y_i(1)-Y_i(0) | G_i=g], \label{eq:pce}
\end{eqnarray}
for $g=AA,CA,NA,DA$. Then the average causal effect of eligibility on dropout is a weighted average of these principal causal effects:
\[\bE[Y_i(1)-Y_i(0)] = \sum_{g=AA,CA,NA,DA} \pi_{g}\tau_{g}, \]
where $\pi_g$ is the proportion of units in principal stratum $g$.

Never-applicants and defiant-applicants never receive a grant, so for them we always observe the outcome in the absence of the grant. By contrast, for always-applicants and compliant-applicants we can observe $Y_i(1)$ for some eligible students who receive a grant and $Y_i(0)$ for some other ineligible students who do not receive a grant. Therefore, always-applicants and compliant-applicants are the only groups where we can learn information about the effect of receiving a grant in this study, and thus the corresponding principal causal effects, $\tau_{AA}$ and $\tau_{CA}$, are the causal estimands of primary interest.

In the standard IV approach to noncompliance \citep{Angrist96, Imbens97}
as well as standard setting of fuzzy RD designs \cite[e.g.,][]{ImbensLemieux08}, data on application status is not utilized, either because it is not available or because it is ignored. Instead, the analysis is based on the principal strata formed by the intermediate variable of grant receipt status. Specifically, there are four principal strata based on the joint potential grant receipt statuses, $R_i=(W_i(0), W_i(1))$: compliers $R_i=(0,1)$, students who would receive the grant if eligible and would not receive the grant if ineligible; always-takers $R_i=(1,1)$, student would receive the grant regardless of eligibility;  never-takers $R_i=(0,0)$, student would not receive the grant regardless of eligibility; and defiers $R_i=(1,0)$, students who would not receive the grant if eligible and would receive the grant  if ineligible. The focus is generally on the causal effect for compliers:
\begin{eqnarray*}
&\tau \equiv \bE[Y_i(1)-Y_i(0)\mid R_i=(0,1)].
\label{tau}
\end{eqnarray*}

We now establish the connection between these two sets of principal strata. The Italian grant assignment rule implies that $W_i(0)=0$ for all $i$, as ineligible units have no access to a grant, and $W_i(1)=0$ if $A_i(1)=0$, as eligible units need to apply for a grant to receive a grant. Therefore, by design, there are no always-takers or defiers, and the remaining principal strata $R$'s can be expressed as unions of principal strata $G$'s: never-takers comprise never-applicants and defiant-applicants, and compliers comprise always-applicants and compliant-applicants. As such, $\tau$ can be rewritten as the weighted average of the causal effects for always-applicants and compliant-applicants:
\begin{eqnarray}
\tau= \bE[Y_i(1)-Y_i(0)\mid G_i \in \{AA,CA\}]= \dfrac{\pi_{AA}\tau_{AA}+ \pi_{CA} \tau_{CA}}{\pi_{AA}+\pi_{CA}}. \label{eq:pce_compliers}
\end{eqnarray}
This illustrates that principal strata defined by the application status leads to a finer partition of the units than principal strata defined by the grant-receipt status. Indeed the standard IV causal estimand---the causal effect for compliers $\tau$ ---provides information on a `marginal' (with respect to application behavior) causal effect. If causal effects are homogeneous, marginalizing over application behavior does not critically affect the evaluation analysis. Conversely, if causal effects are heterogeneous, as we have found in this study, ignoring application behavior represents a loss of useful information with potentially important policy implications. For example, if the grants are found out to have a higher positive effect on always-applicants than on compliant-applicants, then it would be useful and cost-effective to study the characteristics of ineligible applicants and include those into the eligibility rule to allocate additional resources.

The estimands $\tau_{AA}, \tau_{CA}$ and $\tau$ represent effects of eligibility, rather than effects of the receipt of a grant. However, ``the receipt of a grant'' is completely confounded with  ``the eligibility status'': $W(z) = z \times A(z) = z$ for always-applicants and compliant-applicants. To \textit{attribute} these effects to  ``the receipt of a grant'', below we can make an exclusion restriction assumption:
\begin{ass} \label{ER_compliers} (Exclusion Restriction for Compliant-Applicants and Always-Applicants).
For all units with $G_i \in \{AA, CA\}$, or equivalently $R_i=(0,1)$, the effect of eligibility is only through the receipt of the grant.
\end{ass}
Assumption \ref{ER_compliers} attributes the intention-to-treat effect for compliers to the causal effect of the receipt of grant, rather than to its assignment (eligibility). A more formal version of this assumption, which requires double-indexed notations, is given in \cite{ImbensRubin15} (Chapter 23, Assumption 23.4). This type of exclusion restriction is routinely made, often implicitly, in randomized experiments with full compliance \citep{MealliRubin2002, MealliPacini13, ImbensRubin15}.

In real studies, the \textit{sample-average} counterpart of the \textit{population-average} estimands may also be of interest:
\begin{equation}  \label{pce_fs}
\tau_g^{S} \equiv \dfrac{1}{N_g}\sum_{i:G_i =g }[Y_i(1)-Y_i(0)],
\end{equation}
where $g=AA,CA, \{AA,CA\}$ and $N_g$  is the number of units in stratum $g$. Usually the \textit{sample-average} effects can be estimated more precisely than their \textit{population-average} counterparts. The subtle difference between them in Bayesian inference is explained in Section \ref{sec:Bayesian}. More details can be found, for example, in \cite{Rubin78, Imbens97} and \cite{Imbens04}. For simplicity of notation, we do not make the distinction between population-average and sample-average estimands in the methodological discussion, but will present both estimates in the application.

\section{The basis for inference} \label{sec:RDD}
\subsection{Probabilistic treatment assignment mechanism in RD designs} \label{sec:AM}
The complex selection process in Italian university grants system implies that the mechanism governing the receipt of the grant, which depends on both institutional and individual choices, is not ignorable. Below we introduce a probabilistic assignment mechanism underlying the RD design considered here, which is also applicable to general RD settings with minor modifications.

We first define the assignment mechanism, which is a row-exchangeable function that assigns probabilities to all $2^N$ possible $N-$dimensional vectors of assignments $\bZ$, as a row-exchangeable function that assigns probabilities to all possible $N-$dimensional vectors of realizations of the forcing variable, $\bS$, above or below the threshold value, $s_0$. Formally,
\begin{eqnarray} \label{AM}
\lefteqn{\Pr\left(\bZ =\mathbf{z} | \bA(0), \bA(1), \bW(0), \bW(1), \bY(0), \bY(1), \bX \right)}\\
&&=\Pr\left(\bS \in \mathbf{\Lambda}| \bA(0), \bA(1), \bW(0), \bW(1), \bY(0), \bY(1), \bX \right), \nonumber
\end{eqnarray}
where  $\mathbf{z} \in \{0,1\}^N$\ and $\mathbf{\Lambda} \in \Big\{(-\infty, s_0]^N,
(-\infty, s_0]^{N-1}\times (s_0,\infty),   (s_0,\infty) \times (-\infty, s_0]^{N-1},\ldots,$ $(-\infty, s_0] \times (s_0,\infty)^{N-1},  (-\infty, s_0]^{N-1}\times (s_0,\infty), (s_0,\infty)^{N} \Big\}$.
Since $Z$ is a deterministic function of $S$, the assignment mechanism can be formulated with respect to either $Z$ or $S$. Here we prefer $S$ because it is the underlying random variable that describes the reasons for the missing and observed values of potential outcomes: a value of $S$ is assigned, which in turn determines a value for $Z$.

Statistical inference for causal effects requires assumptions on the assignment mechanism. We introduce  assumptions that allow us to describe RD settings as classical randomized experiments around the threshold. The assignment mechanism in Equation \eqref{AM} is a classical randomized experiment if  $(i)$ it is individualistic: 
\begin{eqnarray*}
\lefteqn{\Pr\left(\bS \in \mathbf{\Lambda}| \bA(0), \bA(1), \bW(0), \bW(1), \bY(0), \bY(1), \bX \right)}\\
&&=\prod\limits_{i=1}^{n}\Pr\left(S_i\leq s_0 | A_i(0), A_i(1), W_i(0), W_i(1), Y_i(0), Y_i(1),  \bX_i\right);
\end{eqnarray*}
$(ii)$ it is probabilistic, which implies that for each unit, $i$,  both events $S_i \leq s_0$ and $S_i > s_0$ have \textit{a priori} a non-zero probability of occurring; $(iii)$ it is unconfounded, that is, free of dependence of any potential outcomes; and $(iv)$ it is a known function of its arguments.

The particular assignment rules underlying RD designs suggest that these assumptions are more reasonable for subpopulations of units who have a relatively large probability that the realized values of the forcing variable fall in a neighborhood around the threshold, $s_0$. For these subpopulations, we can reasonably assume that the distribution of the forcing variable is unrelated to observed and unobserved characteristics of students. On the other hand students with a very small (close to zero) or a very large (close to one) probability that $S_i \leq s_0$ are likely systematically different from other students. For example, potential outcomes observed for very rich students, who do not receive any grant, are plausibly different from potential outcomes for poor students with a realized value of $S$ around the threshold, who do not receive a grant, and vice versa. Therefore we focus on subpopulations of students who have a probability that $S_i \leq s_0$ strictly between zero and one, and sufficiently far away from zero and one. The following assumption guarantees that at least one such subpopulation of units exists.
\begin{ass} \label{Ass_overlap}
(Local overlap). Let ${\cal{U}}$ be the random sample (or population) of units in the study.
There exists a subset of units, ${\cal{U}}_{s_0}$, such that for each $i \in {\cal{U}}_{s_0}$, $
\Pr(S_i \leq s_0) > \epsilon$ and $\Pr(S_i > s_0) > \epsilon$  for some sufficiently large $\epsilon >0$.
\end{ass}

Assumption \ref{Ass_overlap} assumes that there exists a subpopulation of units, each of whom has a non-zero probability of being assigned to either treatment levels. This represents a main distinction between our framework and the existing RD literature that often describes RD designs as settings where the overlap assumption is violated. Now within the subpopulation $\Uset$ we can formally introduce a modified SUTVA specific to the RD settings:
\begin{ass} \label{Ass_RD-SUTVA} (Local RD-SUTVA). For each $i \in \Uset$, consider two eligibility statuses
$Z^{'}_i = \one(S_i' \leq s_0)$ and $Z^{''}_i = \one(S^{''}_i \leq s_0)$, with possibly $S_i^{'} \neq S_i^{''}$.  If  $Z^{'}_i = Z^{''}_i$, that is, if either $S^{'}_i \leq s_0$  and $S^{''}_i \leq s_0$, or $S^{'}_i > s_0$  and  $S^{''}_i > s_0$, then $A_i(\bZ^{'})=A_i(\bZ^{''})$, $W_i(\bZ^{'})=W_i(\bZ^{''})$, and $Y_i(\bZ^{'})=Y_i(\bZ^{''})$.
\end{ass}

Local RD-SUTVA rules out interference between units, implying that potential outcomes for a student cannot be affected by the eligibility status of other students. Local RD-SUTVA also assumes that there are no levels of the eligibility status other than zero and one. This component of RD-SUTVA implies that values of the forcing variable leading to the same eligibility status cannot alter potential outcomes for any unit, and thus allows us to avoid defining potential outcomes as functions of the forcing variable. Under the local RD-SUTVA for each unit within $\Uset$ there exist only two potential outcomes for each post-assignment variable, corresponding to the realized value of the forcing variable falling \textit{below} and \textit{above} the threshold, respectively.

Finally, we need to formalize the concept of RD design as local randomized experiment: in a neighborhood of the threshold $s_0$ the forcing variable does not depend on either the potential outcomes or pre-treatment variables. Formally, we have:
\begin{ass} \label{Ass_randomization} (Local randomization). For each  $i \in {\cal{U}}_{s_0}$,
$$
\Pr\left(S_i | A_i(0), A_i(1), W_i(0), W_i(1), Y_i(0), Y_i(1), \bX_i\right) =\Pr\left(S_i  \right).
$$
\end{ass}
Assumption \ref{Ass_randomization}  states that within the subpopulation $\Uset$ a Bernoulli trial has been conducted, with individual assignment probabilities, that is, the individual probabilities of being eligible to receive a grant, depending only on the distribution of the forcing variable: $\Pr(Z_i=1)=\Pr(S_i\leq s_0)$. This assumption is crucial in justifying the key idea underlying any RD design. It implies that the eligibility statuses are randomly assigned in some small neighborhood, $\Uset$, around $s_0$.

Assumption \ref{Ass_randomization} may not always be plausible. For instance, when the forcing variable is a deterministic variable, which conceptually cannot be interpreted as a random variable with a non-degenerate probability distribution (such as time), the underlying design cannot, in general, be interpreted as a local randomized experiment \cite[see Section 6.3 in][pp 347]{LeeLemieux10}.

There are subtle but substantive differences between local RD-SUTVA and local randomization.
Local RD-SUTVA is an exclusion restriction assumption and it is required to make the representation of potential outcomes as functions of the eligibility status adequate. Local randomization is an independence assumption and it is crucial to make inference. RD-SUTVA is different from independence assumptions: it does not imply that  the probability that we observe a value of the forcing variable above or below the threshold does not depend on potential outcomes. RD-SUTVA simply implies that the exposure to assignment level $z$ specifies well-defined potential outcomes, for all unit $i$ and assignment levels $z$. In other words, considering potential outcomes as random variables,  RD-SUTVA does not imply that potential outcomes  have the same distribution for each value of the forcing variable. In order to make the forcing variable independent of potential outcomes, we need to introduce additional assumptions, such as Assumption \ref{Ass_randomization}.

Following Assumption \ref{Ass_overlap}, we can define a local version of the target estimands:
\begin{equation}\label{pce_s0}
\tau_{g,s_0} \equiv \mathbb{E}\left[Y_i(1)-Y_i(0) \mid G_i = g, i \in {\cal{U}}_{s_0} \right],
\end{equation}
for $g=AA,CA, \{AA,CA\}$ and their \textit{finite-sample} counterparts, and we have:
\begin{equation*}\label{tau_frd_s0}
\tau_{\{AA,CA\},s_0} \equiv \tau_{s_0} =\dfrac{\tau_{AA,s_0} \pi_{AA,s0} + \tau_{CA,s_0}\pi_{CA,s0}}{\pi_{AA,s_0}+\pi_{CA,s_0}},
\end{equation*}
where $\pi_{g,s_0}=\Pr(G_i=g| i \in {\cal{U}}_{s_0})$ for $g=AA,CA,NA,DA$, denote the proportion of principal strata in the subpopulation ${\cal{U}}_{s_0}$. A special case of ${\cal{U}}_{s_0}$ contains the subpopulation of units with a realized value of the forcing variable \textit{exactly} equal to the threshold value, $s_0$.

It is worth noting that Assumption \ref{Ass_randomization} implies that
\begin{eqnarray*}
\mathbb{E}\left[Y_i(1)- Y_i(0) \mid  G_i=g, i \in {\cal{U}}_{s_0} \right]   &=& \mathbb{E}\left[Y_i(1)- Y_i(0) \mid Z_i=1, G_i=g, i \in {\cal{U}}_{s_0} \right].
\end{eqnarray*}
Under the allocation rule of the Italian university grants, $Z_i=W_i^{obs}$ for always-applicants and compliant-applicants. Therefore, the local randomization assumption allows the estimands $\tau_{AA,s_0}$, $\tau_{CA,s_0}$, and $\tau_{s_0}$ to be interpreted as causal effects of receiving a grant for subpopulations of students who actually receive a grant, analogous to the notion of average treatment effect for the treated.

\subsection{Two additional assumptions}
The following two assumptions---though not necessary for Bayesian inference---are plausible in our study and can sharpen the inference.

\begin{ass} Monotonicity of Application Status: \label{Ass_mono}
\[A_i(1) \geq A_i(0), \qquad \mbox{for all} ~ i \in {\cal{U}}_{s_0}.\]
\end{ass}

\begin{ass} Stochastic Exclusion Restriction for Never-Applicants: \label{Ass_ER}
\[
\Pr(Y_i(1)|G_i=NA, i \in {\cal{U}}_{s_0})=\Pr(Y_i(0)|G_i=NA, i \in {\cal{U}}_{s_0}).
\]
\end{ass}
Monotonicity rules out the existence of defiant-applicants. The exclusion restriction rules out direct effects of eligibility on dropout  for never-applicants. Never-applicants are students who would never apply for a grant irrespective of their eligibility status. These students would not receive the grant in any case.  Exclusion restriction for never-applicants (Assumption \ref{Ass_ER}) is of very different nature from the exclusion restriction for compliant-applicants and always-applicants (Assumption \ref{ER_compliers}): Assumption \ref{Ass_ER} has implications for inference but not for interpretation, whereas Assumption \ref{ER_compliers} is made solely for interpreting the causal effects of assignment on the outcome attributable to the causal effects of treatment on the outcome. More discussions on the difference can be found in \citet[][Chapter 23]{ImbensRubin15} and \cite{MealliPacini13}.

\subsection{Selection of the subpopulations}  \label{sec:Bandwidth}
An important issue in practice is the selection of the subpopulation $\Uset$ where the RD assumptions hold. There can be a diverse choice of the shape of the subpopulation. In this paper, we limit our choice to symmetric intervals with respect to $s_0$, for convenience and also to match the common practice of RD analysis. Specifically, we make the following assumption:
\begin{ass} \label{Ass_Sneighbor}
There exists $h >0$ such that for each $\epsilon>0$,  $\Pr(s_0 - h \leq S_i \leq s_0+ h) > 1-\epsilon$, for each  $i \in {\cal{U}}_{s_0}$.
\end{ass}
Assumption~\ref{Ass_Sneighbor} allows us to focus on the specific subsets of symmetric intervals among all neighborhoods of different shape around the threshold, $s_0$. Note that Assumptions \ref{Ass_overlap} and  \ref{Ass_Sneighbor}  do not imply that $\Uset$ is unique. They only require that there exists at least one subpopulation, $\Uset$. Consequently, we are not interested in finding the largest $h$, but we only aim at determining plausible values for $h$.

Our approach for selecting bandwidth $h$ exploits the fact that Assumption \ref{Ass_randomization} is a local randomization assumption, in the sense that it holds for a subset of units, but may not hold in general for other units. As such, under Assumption \ref{Ass_randomization}, in the subpopulation $\Uset$, pre-treatment variables should be well balanced in the two subsamples defined by assignment, and thus any test of the null hypothesis of no effect of assignment on pre-treatment covariates should fail to reject the null.

Assessing balance in the observed covariates raises problems of multiple comparisons, which may lead to a much higher than planned type I error if they are ignored \cite[e.g.,][]{BenjaminiHochberg95}. We account for multiplicities using a Bayesian hierarchical mixed model, which provides an explicit method for borrowing information across covariates  \cite[e.g.,][]{BerryBerry04, ScottBerger06}.
Following \cite{BerryBerry04}, we use a mixture for the prior distribution of the eligibility parameters by assigning a point mass on equality of the means of the covariates between eligible and ineligible units. This Bayesian procedure provides a measure of the risk (posterior probability) that a chosen interval around the threshold, $s_0$, defines a subpopulation of units that does not exactly matches any true $\Uset$,  including subjects for which our RD assumptions do not hold. More details are given in Section~\ref{sec:Application}. The idea to exploit balance tests of pre-assignment variables to select a subpopulation of  units is also used in \cite{Cattaneo15}, but their approach aims at selecting the \textit{largest} subpopulation and does not account for multiple comparisons.

Our approach parallels more conventional RD approaches based  on local polynomial regression, which also involve bandwidth selection, but for a very different objective, namely finding an optimal balance between precision and bias at the threshold for local polynomials \cite[e.g.,][]{LudwigMiller07, LeeLemieux10,ImbensKalyanaraman12}, whereas the objective in our framework is to find a subpopulation where our RD assumptions are plausible and the selected subpopulation defines the target population.

\subsection{Mode of inference} \label{sec:InferenceMode}

Once the subpopulation $\Uset$ is chosen, and under the RD assumptions \ref{Ass_overlap}-\ref{Ass_randomization}, one can choose different modes of inference for the target causal estimands, as in the large literature of principal stratification. For example, under the additional Assumptions \ref{Ass_mono} and \ref{Ass_ER}, the average causal effect for compliers,  $\tau_{s_0}$, is non-parametrically point identified and could be estimated using standard moment-based (instrumental variable) methods. But the average causal effects for always-applicants and compliant-applicants, $\tau_{AA,s_0}$ and $\tau_{CA,s_0}$ can be only non-parametrically partially identified \citep{MealliPacini13}. One can also use  likelihood approaches  to parametrically estimate causal effects \citep[e.g.,][]{Frumento12, Mercatanti13}. Randomization-based inference \citep{Fisher25}, as in \cite{Cattaneo15}, could also be adopted.

In this article, we choose the Bayesian approach for inference for the following reasons. First, causal inference in RD designs usually involves complex observational data, with multiple sources of uncertainties, including the missing potential outcomes; the Bayesian approach is particularly useful for accounting for uncertainties and for pooling information from the data in such complex settings. Second, RD analysis usually relies on a sample of units with values of the forcing variable close to a single point, the size of which may be small; Bayesian methods, not relying on asymptotic approximations, are attractive in dealing with small samples.  Third, in the Bayesian paradigm, the missing potential outcomes are treated as random variables, and all inferences are based on the posterior distributions of causal estimands, which are functions of potential outcomes. Thus inference about finite-sample and super-population estimands can be drawn using the same inferential procedures. Finally, pre-treatment variables can be easily incorporated in the Bayesian approach, which may improve efficiency of the analysis, i.e., reduce posterior variability.

\section{Bayesian inference} \label{sec:Bayesian}
Our development of the Bayesian approach builds on the seminal works of \cite{Rubin78} and \cite{Imbens97}. Below we give a brief outline for conduction principal stratification analysis using a Bayesian approach; the readers may refer  to the existing literature for more details \citep[e.g.,][]{Elliott10,Schwartz11,MLM13}. Throughout the discussion, we use $p(\cdot|\cdot)$ and $\btheta_{\cdot|\cdot}$ to denote generic conditional distributions and the corresponding parameters, respectively.

Nine quantities are associated with each unit: $Y_i(0)$, $Y_i(1)$, $W_i(0)$, $W_i(1)$, $A_i(0)$, $A_i(1)$, $\bX_i$, $Z_i$, $S_i$. Among these, $S_i$ completely determines $Z_i$; the principal stratum $G_i=(A_i(0), A_i(1))$ and $S_i$ completely determine $(W_i(0),W_i(1))$. Therefore, inference for causal effects involves only $Y_i(0)$, $Y_i(1)$, $A_i(0)$, $A_i(1)$, $\bX_i$, $S_i$, of which four are observed: $S_i$, $\bX_i$, $\Aobs=A_i(Z_i)$, $\Yobs=Y_i(Z_i)$, and two are unobserved: $\Amis=A_i(1-Z_i)$, $\Ymis=Y_i(1-Z_i)$.

Bayesian inference considers the observed values to be realizations of random variables and the unobserved values to be unobserved random variables.  Let $p(\bY(0), \bY(1), \bA(0), \bA(1),$ $\bX, \bS; \Uset)$ denote the joint probability density function of these random variables for all units in $\Uset$. We assume this distribution is unit-exchangeable, that is, it is invariant under a permutation of the unit indices. Then, with essentially no loss of generality, by appealing to de Finetti's theorem \citep{Finetti63}, we can assume that there exists an unknown parameter vector $\btheta$, which is itself a random variable having a known prior distribution $p(\btheta)$ such that:
\begin{eqnarray*}
p\left(\bY(0), \bY(1), \bA(0), \bA(1), \bX, \bS; \Uset \right)=\int \prod_{i \in \Uset} p\left(Y_i(0), Y_i(1), A_i(0), A_i(1), \bX_i, S_i | \btheta \right)p(\btheta) d\,\btheta.
\end{eqnarray*}
Bayesian inference of the causal estimands, which are functions of $Y_i(z)$'s and $A_i(z)$'s, centers around deriving the posterior distribution for the parameter vector of their distribution, denoted by $\btheta_{Y,G}$. Under Assumption \ref{Ass_randomization},  and assuming the parameters governing the distributions of the covariates, the forcing variable,  and the potential outcomes are \textit{a priori} distinct and independent, the posterior distribution of $\btheta_{Y,G}$ can be written as follows:
\begin{eqnarray} \label{eq:decompose2}
\lefteqn{p\left(\btheta_{Y,G} | \bY^{obs}, \bA^{obs}, \bX, \bS; {\cal{U}}_{s_0}\right) \propto p(\btheta_{Y|G})\times p(\btheta_{G})\times }\\
 && \prod\limits_{i\in \Uset}	
\Big[ \int \!\!\int  p\left(Y_i(0), Y_i(1) | G_i, \bX_i; \btheta_{Y|G} \right) p\left(G_i |  \bX_i; \btheta_{G} \right) \ d\,Y_i^{mis} d\, A_i^{mis}\Big]. \nonumber
\end{eqnarray}
The above decomposition suggests that two models need to be specified for model-based inference: (1) the model for potential outcomes conditional on principal strata and covariates, and (2) the model for principal strata conditional on covariates, as well as the prior distribution for the parameters, $p(\btheta_{Y,G})$, with $\btheta_{Y,G}=(\btheta_{G},\btheta_{Y|G})$.

Let $\pi_{i,g}=\Pr( G_i=g | \bX_i; \btheta_{G})$ and $f_{i,gz}=p(Y_i(z) | G_i=g, \bX_i; \btheta_{Y|G})$. Then the posterior distribution of $\btheta_{Y,G}$ given the observed data can be written as follows:
\begin{eqnarray}\label{eq:likelihood}
\lefteqn{p\left(\btheta_{Y,G} | \bY^{obs}, \bA^{obs}, \bX, \bS; {\cal{U}}_{s_0}\right)} \\
&\propto& p(\btheta_{Y,G}) \times
\prod\limits_{i \in {\cal{U}}_{s_0}: S_i > s_0, A_i^{obs}=0}
\left(\pi_{i,CA}f_{i,CA,0} + \pi_{i,NA}f_{i,NA}\right) \times \prod\limits_{i \in {\cal{U}}_{s_0}: S_i > s_0, A_i^{obs}=1}\pi_{i,AA}f_{i,AA,0} \nonumber\\
&& \times \prod\limits_{i \in {\cal{U}}_{s_0}: S_i \leq s_0, A_i^{obs}=0}\pi_{i,NA}f_{i,NA}
\times \prod\limits_{i \in {\cal{U}}_{s_0}: S_i \leq s_0, A_i^{obs}=1}
\left(\pi_{i,AA}f_{i,AA,1} + \pi_{i,CA}f_{i,CA,1}\right), \nonumber
\end{eqnarray}
where $f_{i,NA}= f_{i,NA,0}=f_{i,NA,1}$ by the exclusion restriction (Assumption \ref{Ass_ER}).
The likelihood function, specified by the four products, does not depend on the association between the potential outcomes $Y_i(0)$ and $Y_i(1)$. Therefore the posterior distribution of the association parameters equal their prior distribution as long as the association parameters are \textit{a priori} independent of the other parameters, as we assume henceforth. The population-average causal estimands $\tau_{{AA,s_0}}$, $\tau_{{CA,s_0}}$, and $\tau_{{s_0}}$ are functions of the parameter vector $\btheta_{Y,G}$, which is free of the association parameters, therefore inference for them does not involve the association parameters \citep[also see discussion in][]{Imbens97}. Inference for sample-average causal estimands for the units in the study, on the other hand, do generally involve the association parameters. In our application inference for sample-average causal estimands is drawn under the assumption that for each unit $i$, potential outcomes, $Y_i(0)$ and $Y_i(1)$, are independent conditional on $\bX_i$ and $\btheta$.

Posterior inference of $\btheta_{Y,G}$ can be obtained using Gibbs sampling with a data augmentation step to impute the missing $A_i^{mis}$,   iteratively drawing from the two posterior predictive distributions, $p\left(\btheta_{Y,G} | \bY^{obs}, \bA^{obs}, \bA^{mis}, \bX, \bS; {\cal{U}}_{s_0}\right)$ and $p\left(\bA^{mis} | \bY^{obs}, \bA^{obs}, \bX, \bS, \btheta_{Y,S}; {\cal{U}}_{s_0}\right)$.

Specification of $\pi_{i,g}, f_{i,gz}$ and corresponding prior to posterior computation depends on the specific application. Details of the models and computation in our application will be provided in Section \ref{sec:Application}. As a general guideline, we recommend to specify $\pi_{i,g}$ and $f_{i,gz}$ conditional on both covariates $\bX$ and the forcing variable $S$, even though Equation \eqref{eq:decompose2} suggests conditioning on $S$ is not required. Indeed, if the true subpopulations ${\cal{U}}_{s_0}$ were known, in theory, we would not need to adjust for $S$, because local randomization guarantees that for units in ${\cal{U}}_{s_0}$ values of the forcing variable falling above or below the threshold are independent of the potential outcomes. However, in practice, the true subpopulations ${\cal{U}}_{s_0}$  are usually unknown and once  a subpopulation has been selected, that is, once a value for $h$,  say $h^\ast$, has been chosen, there may be some units with a realized value of $S$ between $s_0-h^\ast$ and $s_0+h^\ast$ who do not belong to ${\cal{U}}_{s_0}$. For these units there may be a relationship between the forcing variable and potential outcomes, and these potential dependences need to be modeled. Specifically, systematic differences in the forcing variable $S$ that, by definition, occur between eligible and ineligible units, may affect inference in the presence of students who do not belong to ${\cal{U}}_{s_0}$.

\section{Evaluation of Italian university grants} \label{sec:Application}
\subsection{Data} We apply the proposed method to the data from the cohort of first-year students enrolled in 2004 to 2006 at University of Pisa and University of Florence. For each student, information on grant application status ($\Aobs$), grant receipt status ($\Wobs$) at the beginning of the academic year, dropout status at the end of the academic year, and covariates ($\bX_i$) is obtained  from  ministry of education and university administrative records. The forcing variable $S$ is a combined economic measure of each student, calculated from one's income tax return and property adjusted for family size based on a formula that is typically not fully known to the students. In all three years, the threshold of eligibility is the combined economic measure of a student below $15\,000$ euros. Thus, the eligibility status ($Z$) is also observed. Typically, students need support from fiscal experts to compute their value of $S$, and the income revenue authority conducts random inspections to verify that the official tax return were reported. These factors make extremely difficult, if not impossible, for students or students' families to  manipulate the value of $S$ in order to end up on the right side of the threshold. Therefore we argue that the local randomization assumption is reasonable here. Ineligible students apply either usually because they are not fully aware of their eligibility status, or because they hope that their application will be still considered because of extra funding or other considerations.

Covariates include sex, high school grade, high school type (4 categories), major in university (6 categories), indicator of year of enrollment (2004, 2005, 2006) and indicator of university (Pisa vs. Florence). Note that the data only include students who had a high school grade of at least $70/100$ and applied either for a grant or for a reduction of tuition fee.
Summary statistics of important variables for the students with the combined economic measure $S$ within 1000 euros of the threshold  are given in Table \ref{tab:summary}. An unadjusted comparison would suggest that the applicants have higher high-school grades, which is an important indicator of a student's academic performance, but also higher dropout rate regardless of their eligibility status.

\begin{figure}[h]
\vspace{6pc}
 \caption{Application rate (a) and dropout rate (b) as a function of the forcing variable for the first year students in Universities of Florence and Pisa in 2004$-$2006. The smoothed lines are estimated using polynomial logistic regression models (of order 3) on each side of the threshold separately; each point are calculated from the units within a binwidth of 50 euros.}
 \label{fig:AYvsS}
\centering
\subfigure[Application rate] 
{\label{fig:AvsS}
    \includegraphics[angle=0,width=2.5in,height=2.5in]{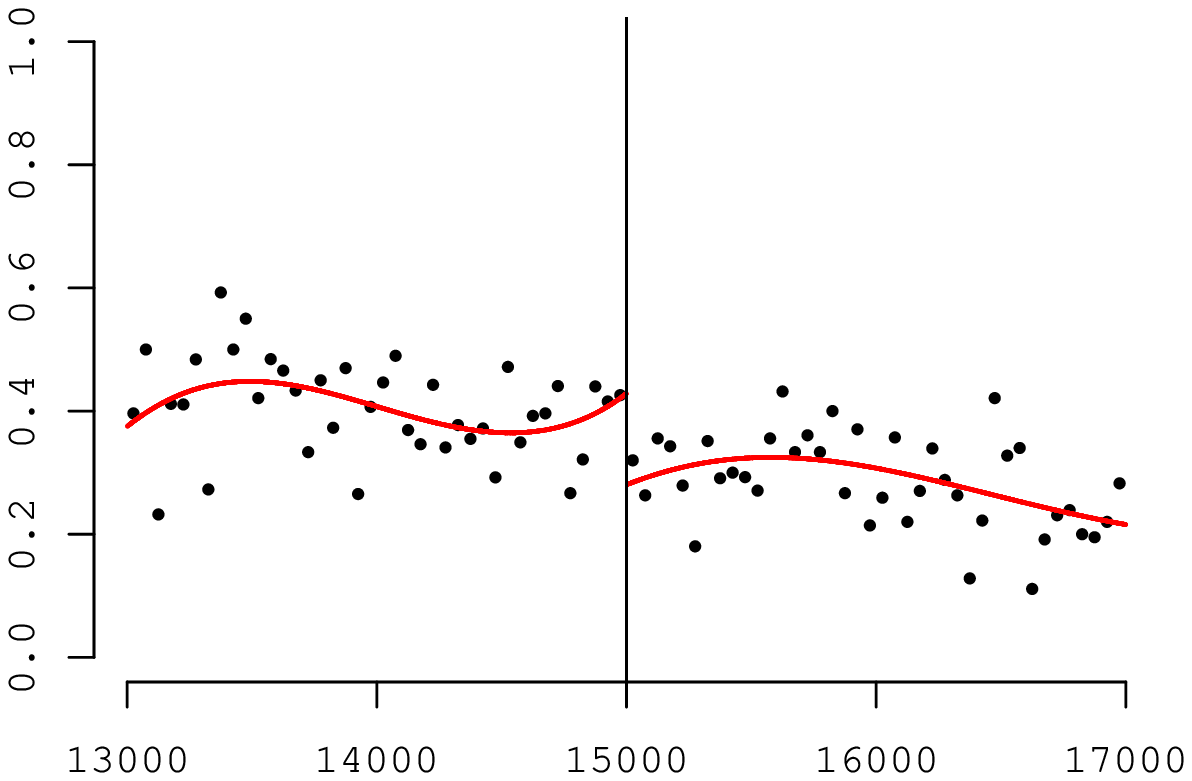}
}
\subfigure[Dropout rate. ~~\tiny{$\square \,$} \textcolor{blue}{\textemdash $\!\!$ \textemdash} \ \  Non-Applicants, ~~  \tiny{\textbullet} \textcolor{red}{\textemdash $\!\!$ \textemdash} \ \ Applicants]
{
    \label{fig:YvsS}
    \includegraphics[angle=0,width=2.5in,height=2.5in]{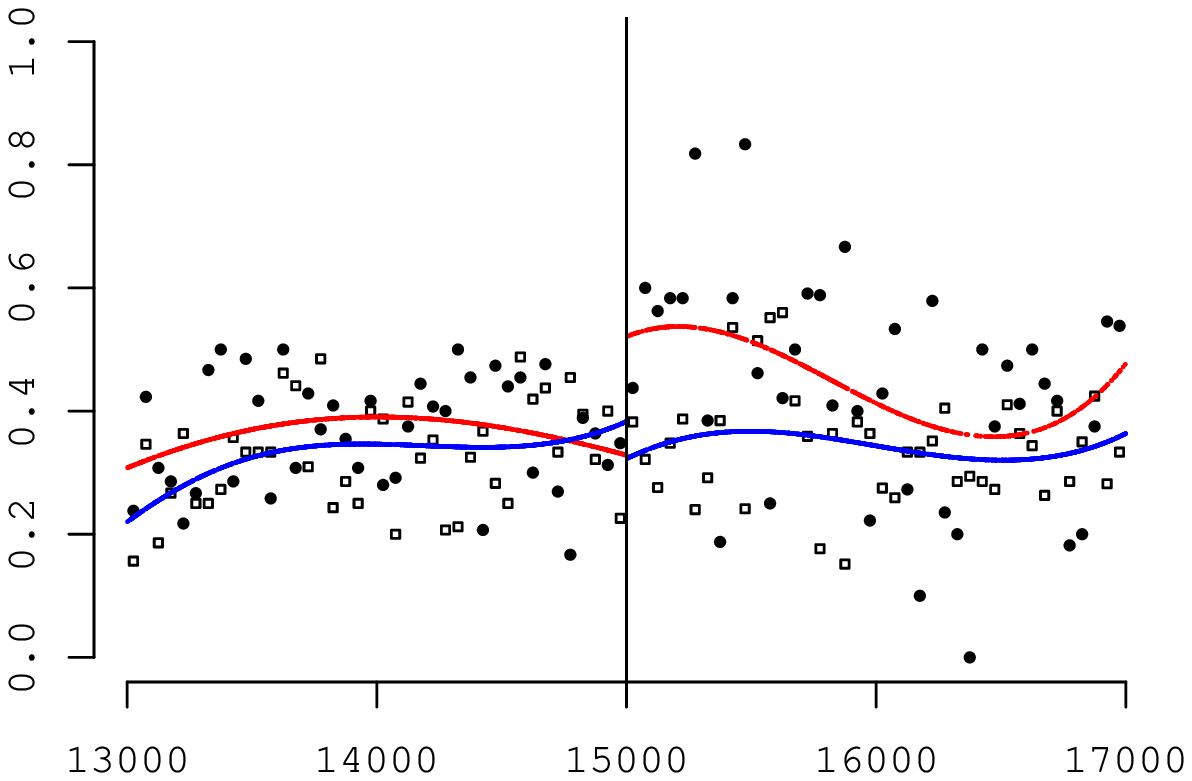}
}
\end{figure}

Application rate and dropout rate as a function of $S$ among the students are given in Figure \ref{fig:AYvsS}. The overall dropout rate is high, consistently between 30\% to 50\% regardless of the economic measure.  From the fitted lines using local logistic polynomial models with order 3 on the two sides of the threshold, discontinuity is clearly visible in both application rate and dropout rate at the threshold. As the economic measure increases, application rate steadily decreases, while the trend in dropout rate has a concave change at the threshold, increasing on the left of the threshold and decreasing on the right.

\begin{table}[h]
\centering
\caption{Summary statistics of the first-year students enrolled in $2004-2006$ at Universities of Pisa
and Florence, for the students with $S_i \in (14\,000, 16\,000)$ euros (i.e., $h=1\,000, s_0=15\,000$).}
\label{tab:summary}
\begin{tabular}{l c rr c rr}
  \hline
        &&\multicolumn{2}{c}{$Z=0$}  && \multicolumn{2}{c}{$Z=1$}\\
\cline{3-4} \cline{6-7}
Variable&&{$A^{obs}=0$}  &{$A^{obs}=1$} &&{$A^{obs}=0$}  &{$A^{obs}=1$}\\
  \hline
   Sample Size  &&  $657$         &  $304$          && $703$           &$444$    \\
   Dropout      &&  $0.36$        &  $0.50$         && $0.35$          & $0.36$   \\
  $S$ (euros) 	&&  $15\,495$		&	$15\,509$       &&	$14\,504$   	&	$14\,499$   		\\
  Female        &&	$0.59$	    &	$0.61$		&&	$0.60$		&	$0.55$			\\
  HS Grade 	    &&	$80.80$		&	$84.35$		&&	$80.17$		&	$84.47$			\\
  University (Pisa) 	&&	$0.37$		&	$0.51$	    &&	$0.37$		&	$0.51$			 \\
\hline
\end{tabular}
\end{table}

\subsection{Selection of the subpopulation} We apply the Bayesian approach to multiple testing  discussed in Section \ref{sec:Bandwidth} to find subpopulations of units where our RD assumptions hold. Specifically we use a hierarchical Bayesian model for assessing the balance of the covariates between eligibility groups. We specify probit models for binary variables; conditional probit models for categorical variables and Gaussian models for continuous variables. Formally, we assume that
$X_j \sim N(\gamma_{0j} + \gamma_{1j} Z_i, \sigma^2_j)$ if $X_j$ is continuous,  and
$\Pr(X_{ij}=1) = Pr(X^\ast_{ij}>0)$ with $X^\ast_{ij} \sim N\left(\gamma_{0j} + \gamma_{1j} Z_i,1\right)$,
if $X_j$ is binary. If $X_j$ is a categorical variable taking on $K$ values we assume that $\Pr(X_{ij}=1) = \Pr\left(X^{\ast (1)}_{ij}\leq 0\right)$, and
$\Pr(X_{ij}=k) = \Pr\left(\cap_{\ell=1}^{k-1} \{ X^{\ast (\ell)}_{ij} > 0\} \cap X^{\ast (k)}_{ij}\leq 0\right)$ for $k =2, \ldots, K-1$, where $X^{\ast (k)}_{ij} \sim N\left(\gamma^{(k)}_{0j} + \gamma^{(k)}_{1j} Z_i, 1\right)$, $k=1, \ldots, K-1$,  independently. Let $\bgamma_{0j} = \left(\gamma^{(1)}_{0j},\ldots,
\gamma^{(K-1)}_{0j}\right)'$ and $\bgamma_{1j} = \left(\gamma^{(1)}_{1j},\ldots,
\gamma^{(K-1)}_{1j}\right)'$.

We specify the following prior  distributions for the model parameters. The variances of the continuous variables have an inverse-Gamma distribution: $ \sigma_j^2 \sim IG(\underline{a}, \underline{b})$. The $\gamma_0$'s have Gaussian prior distributions: for continuous and binary variables, $ \gamma_{0j} \sim N(\mu_{\gamma_0}, \sigma^2_{\gamma_0})$, and for categorical variables, $\bgamma_{0j} \sim N(\mu_{\gamma_0}\boldsymbol{u}_{K-1}, \sigma^2_{\gamma_0}\boldsymbol{I}_{K-1})$  with $\boldsymbol{u}_{K-1}$ and $\boldsymbol{I}_{K-1}$ being the $K-1$-dimensional vector of ones and the identity matrix of order $K-1$, respectively.
Further, for continuous and binary variables, parameters $\gamma_{1j}$ are the difference between means/proportions for eligible and ineligible units. If $\gamma_{1j}=0$ then $X_j$ has the same distribution for eligible and ineligible units. For a categorical variable taking on $K$ values, the proportion of units in each category is the same for eligible and ineligible units if and only if $\gamma^{(k)}_{1j}=0$ for each $k=1, \ldots, K-1$. We assign positive probability to these possibilities using the following mixture prior
distributions:
$$
\gamma_{1j} \sim \pi_{\gamma_1} \delta_{0}(\gamma_{1j}) + (1-\pi_{\gamma_1})N(\mu_{\gamma_1}, \sigma^2_{\gamma_1})
$$
and
$$
\bgamma_{1j} \sim \prod_{k=1}^{K-1}\left[\pi_{\gamma_1} \delta_{0}(\gamma^{(k)}_{1j}) + (1-\pi_{\gamma_{1}})N(\mu_{\gamma_1}, \sigma^2_{\gamma_1})\right],
$$
where $\delta_{0}(\cdot)$ is the Dirac delta distribution.

For the hyperparameters, we assign the following prior distributions: $\mu_{\gamma_0} \sim  N(\underline{\mu}_{\gamma_0}, \underline{\sigma}^2_{\gamma_0})$; $\sigma^2_{\gamma_0} \sim IG(\underline{a}_{\gamma_0}, \underline{b}_{\gamma_0})$;
$\mu_{\gamma_1} \sim  N(\underline{\mu}_{\gamma_1}, \underline{\sigma}^2_{\gamma_1})$; $\sigma^2_{\gamma_1} \sim IG(\underline{a}_{\gamma_1}, \underline{b}_{\gamma_1})$; and $\pi_{\gamma_1} \sim Beta(\underline{a}_{\pi}, \underline{b}_{\pi})$.

We implement the Bayesian model for assessing the balance of covariates on the two sides of the threshold for various subpopulations defined by different values of $h$.  Details on the Monte Carlo Markov Chain (MCMC) for the posterior computation are relegated to
\ref{websupp}. Table \ref{tab:balance} shows the posterior probabilities that the covariates have the same distribution between eligible and ineligible students for the subpopulations defined by
$h=250$, $500$, $750$, $1\,000$, $1\,500$, $2\,000$, $2\,500$, $3\,000$, $4\,000$, $5\,000$.
These values show that the probability of the pre-assignment variables being well balanced is high for subpopulations defined by values of $h$ strictly lower than $1\,500$: the vast majority of these probabilities are larger than or close to 0.8. Note that the probabilities are in general lower among the covariates of ``major in university'', suggesting these covariates may not be as balanced as other covariates. Nonetheless, nearly all these probabilities are still higher 0.6 with a single lowest probability being 0.565 (Tech major in university). For larger subpopulations some covariates, such as the ``indicator of university,''  are clearly unbalanced.

Given that the risk that a chosen interval around the threshold defines a subpopulation that includes units not belonging to the target subpopulation, $\Uset$, is not zero, in order to account for the presence of these units, we conduct the subsequent analyses conditioning on both covariates and the realized values on the forcing variable. Also we evaluate the robustness of our results conducting analyses using various values of $h$ ($h=500, 1\,000, 1\, 500$)
\begin{sidewaystable}[h]
\centering
\caption{Posterior probabilities that the covariates have the same distribution between eligible and ineligible students for various subpopulation} \label{tab:balance}
\begin{footnotesize}
\begin{tabular}{l cccccccccc} \hline
           &  $h$=250    & $h$=500     & $h$=750     & $h$=1\,000    & $h$=1\,500   & $h$=2\,000   & $h$=2\,500   &$h$=3\,000   & $h$=4\,000  & $h$=5\,000\\
{Variable} & ($n$=528)   &($n$=1\,042) & ($n$=1\,577)& ($n$=2\,108)  & ($n$=3\,166) & ($n$=4\,197) & ($n$=5\,159) &($n$=6\,113) &  ($n$=8\,061)  & ($n$=9\,846) \\
\hline
{{Sex}}        & .955 & .950 & .960 & .962 & .977 & .970 & .991 & .960 & .968 & .797 \\
\multicolumn{11}{l}{{High  School Type (Baseline: Other)}}	\\
\,\,\,{Humanity}    & .951 & .952 & .949 & .955 & .979 & .970 & .965 & .986 & .953 & .962 \\
\,\,\,{Science}  	& .894 & .905 & .926 & .927 & .951 & .889 & .916 & .926 & .045 & .000 \\
\,\,\,{Tech}        & .790 & .807 & .790 & .808 & .819 & .619 & .751 & .793 & .003 & .000 \\
{HS Grade}   	    & .955 & .958 & .972 & .978 & .971 & .981 & .987 & .990 & .984 & .986 \\
\multicolumn{11}{l}{{Year (Baseline: 2004)}}								\\
\,\,\, 2005   & .932 & .964 & .954 & .926 & .973 & .977 & .976 & .983 & .861 & .918 \\
\,\,\, 2006	  & .883 & .918 & .914 & .909 & .959 & .934 & .952 & .970 & .807 & .884 \\
{{University (Pisa)}}		& .950 & .916 & .971 & .983 & .686 & .097 & .225 & .300 & .082 & .000 \\
\multicolumn{11}{l}{{Major in University (Baseline: Other)}}\\													
\,\,\,{Humanity} 	& .946 & .899 & .689 & .797 & .798 & .932 & .958 & .990 & .964 & .946 \\
\,\,\,{Science} 	& .894 & .857 & .660 & .751 & .783 & .901 & .929 & .966 & .911 & .913 \\
\,\,\,{Social Science}   & .798 & .821 & .624 & .713 & .758 & .864 & .913 & .953 & .878 & .858 \\
\,\,\,{Bio-Med} & .728 & .776 & .604 & .677 & .736 & .837 & .889 & .926 & .839 & .832 \\
\,\,\,{Tech} & .632 & .634 & .565 & .624 & .699 & .794 & .863 & .876 & .719 & .453 \\ 	
\hline
\end{tabular}
\end{footnotesize}
\end{sidewaystable}

\subsection{Parametric models} For the units within the selected subpopulation $\Uset$, we assume parametric models for the outcome ($f_{gz}$) and principal strata ($\pi_{g}$). Alternative models, such as Student-$t$ models \citep{ChibJacobi11} and Bayesian nonparametric models \citep{Schwartz11}, can  be considered. Note that although we are using parametric models, identification does not rely on parametric assumptions. The model for the principal strata of application consists of two conditional probit models:
\begin{eqnarray*}
\pi_{i,AA} &=& \Pr(G^\ast_i(AA) \leq 0), \\
\pi_{i,NA} &=& \Pr(G^\ast_i(AA) > 0 \, \mathrm{and}  \, G^\ast_i(NA) \leq 0),\\
\pi_{i,CA} &=& 1- \pi_{i,AA}-\pi_{i,NA},
\end{eqnarray*}
where
\[
G^\ast_i(AA) = \alpha_{AA,0} + \alpha_{AA}^{(S)}S^{\ast}_i + \mathbf{X}_i'\balpha_{AA}^{(X)} + \epsilon_{AA,i},
\qquad
G^\ast_i(NA) = \alpha_{NA,0} +  \alpha_{NA}^{(S)}S^{\ast}_i +\mathbf{X}_i'\balpha_{NA}^{(X)} + \epsilon_{NA,i},
\]
with $\epsilon_{AA,i} \sim N(0,1)$, $\epsilon_{NA,i} \sim N(0,1)$ independently, and $S^{\ast}_i = (S_i-s_0)/1000$.

Dropout, the primary outcome in our application, is binary. Therefore, we assume the following generalized linear outcome model with a probit link \citep{AlbertChib93}:
\[
\Pr(Y_i(z) =1 | G_i=g,  S_i, \mathbf{X}_i) = \Phi\left(\beta_{0,g,z} +
\beta^{(S)}_{g,z}S^{\ast}_i  + \mathbf{X}_i'\bbeta^{(X)}_{g,z}\right).
\]
We impose prior equality of the slope coefficients in the outcome regressions: $\bbeta^{(X)}_{g,z} \equiv \bbeta^{(X)}$ for $g=AA, CA, NA$ and $z=0,1$.

Define $\balpha_{g} = [\alpha_{g0}, \alpha_{g}^{(S)}, \balpha_{g}^{(X)}]'$, $g=AA,NA$, and $\bbeta_{g,z} = [\beta_{0,g,z}, \beta_{g,z}^{(S)}]'$, $g=AA, CA, NA$; $z=0,1$. By Assumption \ref{Ass_ER}, $\bbeta_{NA,0}=\bbeta_{NA,1}$. We assume that parameters are \textit{a priori} independent and use multivariate normal prior distributions:
\[
\balpha_{g} \sim N\left(\underline{\bmu}_{\alpha_g}; \underline{\sigma}^2_{\alpha_g} \boldsymbol{I} \right),
\quad
\bbeta_{g,z} \sim N\left(\underline{\bmu}_{\beta_{g,z}};\underline{\sigma}^2_{\beta_{g,z}} \boldsymbol{I} \right),
\quad
\bbeta^{(X)} \sim N\left(\underline{\bmu}_{\beta}; \underline{\sigma}^2_{\beta} \boldsymbol{I} \right)
\]
where $\boldsymbol{I}$ is the identity matrix. We specify flat priors setting the hyper-parameters as follows: setting $\underline{\bmu}_{\alpha_g}$, $\underline{\bmu}_{\beta_{g,z}}$, $\underline{\bmu}_{\beta}$ to be null vectors; and setting large prior variances $\underline{\sigma}^2_{\alpha_g}=10$, $\underline{\sigma}^2_{\beta_{g,z}}=10$,  $\underline{\sigma}^2_{\beta}=10$ for $g=AA, CA, NA$; $z=0,1$.
\medskip

\subsection{Posterior computation} Details of the MCMC algorithm for the posterior computation based on the outline in Section \ref{sec:Bayesian} are given in \ref{websupp}. Upon obtaining the posterior draws of the parameters, we calculate three estimates for each causal estimand: population-average effect \emph{within} $\Uset$ and \emph{at} $s_0$, and sample-average effect within $\Uset$. The population-average effects within $\Uset$ are calculated averaging the model-based dropout proportions over the empirical distribution of the pre-assignment variables and the forcing variable:
 \[
 \dfrac{\sum_{i \in \Uset} \pi_{i,g}   \Phi\left(\beta_{0,g,1} +
 \beta^{(S)}_{g,1}S^{\ast}_i  + \mathbf{X}_i'\bbeta^{(X)}\right)}{\sum_{i \in \Uset} \pi_{i,g}} - \dfrac{\sum_{i \in \Uset} \pi_{i,g}   \Phi\left(\beta_{0,g,0} +
  \beta^{(S)}_{g,0}S^{\ast}_i  + \mathbf{X}_i'\bbeta^{(X)}\right)}{\sum_{i \in \Uset} \pi_{i,g}},
 \]
for $g = AA,CA, \{AA,CA\}$. The population-average effects at $s_0$ are calculated in a similar way setting $S^{\ast}_i=0$ (i.e., $S_i=s_0$) for each $i$. To obtain the sample-average estimates, we compute the posterior predictive distributions of the potential outcomes for each student $i$ in $\Uset$, based on which the sample average is calculated.

\subsection{Results}
We conducted Bayesian analysis using $h=500, 1\,000, 1\,500$. Posterior inference is  based on $5\,000$ draws from the posterior distributions simulated using single chains, which were run for $125\,000$ iterations. To assess convergence of iterative simulation methods, we calculated the Cramer-von-Mises statistic to test the null hypothesis that the sampled values come from a stationary distribution and visual inspected  the trace-plots of the causal parameters (functions of model parameters).  We also run multiple MCMC chains with different starting for each $h$ to evaluate the mixing of the chains using the Gelman-Rubin statistic \citep{GelmanRubin92}. The results
provided no evidence against convergence\footnote{We also conducted Bayesian analysis using alternative models with different order polynomials in $S$ as well as  models conditioning only on $S$ (without using the pretreatment variables) and null models, conditioning on neither $S$ nor the pre-treatment covariates. Consistently to results found in \cite{MealliRampichini12}, higher order polynomials do not lead to substantial inferential benefits, and posterior distributions of the causal effects of interest did not substantially change  with the alternative models, so here we only show the results based on models conditioning on both $S$ and the pre-treatment covariates.}.

Table \ref{tab:effect} shows posterior medians and 95\% credible intervals for the principal strata proportions under monotonicity and  for the causal parameters $\tau_{AA,s_0}, \tau_{CA,s_0}, \tau_{s_0}$, for bandwidths ranging from $500$ to $1500$ euros. The results are robust across different bandwidths.  The estimated proportions of the principal strata are very similar across different $h$: there are more than 61\% never-applicants, more than 32\% always-applicants and less than 6.5\% compliant-applicants.  The three estimates for the same causal parameter are also similar. The posterior distributions of the causal effect for always-applicants, $\tau_{AA,s_0}$, and the union of always-applicants and compliant-applicants, $\tau_{s_0}$, are centered on negative values, and the 95\% credible intervals do not cover 0, irrespective of the choice of the bandwidth.

For instance, consider the finite-sample causal effects for the subpopulation within $h=1\,000$ euros around the threshold (middle block of columns in Table~\ref{tab:effect}). The estimated $\tau_{s_0}$ suggests a $13.9\%$ (95\% CI: $(3.4\%; 24.7\%)$) reduction in dropout rate for the students who receive the grants. The estimated $\tau_{AA,s_0}$ suggests an even stronger positive effect among the always-applicants: a $16.1\%$ (95\% CI: $(5\%; 27\%)$) reduction in dropout rate. In fact, $\tau_{s_0}$, which is a weighted average of the effects for always-applicants and compliant-applicants, appears to be diluted by the somewhat surprising small effect among the compliant-applicant. However, the data do not seem to contain much information on compliant-applicants (the estimated proportion of compliant-applicants is very small, less than 5\%), and the effects were estimated with large uncertainties.

These results suggest that the current Italian university grants are effective in reducing dropout from universities among students from families with annual economic measure around 15\,000 euros. Our analysis also reveals some additional information for policy making. Specifically, always-applicants and compliant-applicants are found to be heterogeneous with respect to the effect of the grants. The causal effect for compliers, $\tau_{s_0}$, usually estimated in a standard IV analysis that ignores the application information, is attenuated by the  small (and negative) effect estimated for the small proportion of compliant-applicants. From a cost-effective perspective, it appears more beneficial for education administrations to lower the eligibility criteria (i.e., decrease the threshold $s_0$) to allow more applicants to get the grant, than to increase the amount of the grant to awardees. The combination of low percentage of compliant-applicants and high percentage of always-applicants suggests that most students with the economic measure being around the threshold who intend to apply for the grants would apply irrespective of their eligibility. From a policy perspective, this implies that educational administrations should better explain the rule of eligibility to potential applicants to discourage ineligible students from applying, and thus reduce unnecessary efforts from these students and the administration, for processing these applications.

\begin{table}[h]
\centering
\caption{Posterior median and 95\% credible intervals of principal strata proportion and super-population and finite-sample causal effects on dropout for always-applicants ($\tau_{AA,s_0}$), compliant-applicants ($\tau_{CA,s_0}$), and their union ($\tau_{s_0}$), for the subpopulation within different bandwidths $h$ around the threshold.}
\vspace{-0.75cm}
\label{tab:effect}
{\small{
\[
\begin{array}{lcrrcrrcrr}
\hline
&\!\!\!&\multicolumn{2}{c}{\hbox{Population-average}}&\!\!\!&\multicolumn{2}{c}{\hbox{Sample-average}}&\!\!\!&\multicolumn{2}{c}{\hbox{Population-average at }  s_0}\\
\cline{3-4} \cline{6-7} \cline{9-10}
h &\!\!\!& \hbox{Median} & 95\% \hbox{ CI }&\!\!\!& \hbox{Median} & 95\% \hbox{ CI } &\!\!\!& \hbox{Median} & 95\% \hbox{ CI } \\
\hline
\multicolumn{10}{l}{h=500}\\
\Pr(G_i=AA)    &\!\!\!& .323  & (.294; .355)   &\!\!\!& .322 & (.309; .336)    &\!\!\!& .320  & (.291; .352) \\
\Pr(G_i=CA)    &\!\!\!& .060  & (.031; .105)   &\!\!\!& .041 & (.021; .090)    &\!\!\!& .058  & (.030; .094) \\
\Pr(G_i=NA)    &\!\!\!& .616  & (.570; .650)   &\!\!\!& .637 & (.590; .651)    &\!\!\!& .621  & (.583; .654) \\
\tau_{AA,s_0} &\!\!\!& -.153 & (-.313; -.030) &\!\!\!& -.152 & (-.307; -.038) &\!\!\!& -.154 & (-.298; -.030) \\
\tau_{CA,s_0} &\!\!\!& .045  & (-.170; .497)  &\!\!\!& .074  & (-.256; .545)  &\!\!\!& .039  & (-.169; .474) \\
\tau_{s_0}   &\!\!\!& -.116 & (-.253; -.005) &\!\!\!& -.120 & (-.265; -.009) &\!\!\!& -.120 & (-.245; -.012) \\
\hline
\vspace{-0.4cm}\\
\multicolumn{10}{l}{h=1\,000}\\
\Pr(G_i=AA)    &\!\!\!& .336 & (.312;  .365)  &\!\!\!& .333  & (.318; .354) &\!\!\!& .335 & (.311; .363) \\
\Pr(G_i=CA)   &\!\!\!& .043 & (.002;  .086)  &\!\!\!& .027  & (.002;   .075) &\!\!\!& .043 & (.001; .075) \\
\Pr(G_i=NA)   &\!\!\!& .623 & (.584;  .652)  &\!\!\!& .640  & (.599;   .645) &\!\!\!& .625 & (.594; .656) \\
\tau_{AA,s_0}  &\!\!\!&-.161 &(-.273; -.052)  &\!\!\!& -.161 & (-.270;  -.057) &\!\!\!& -.154 & (-.259; -.052) \\
\tau_{CA,s_0}  &\!\!\!& .028 &(-.745;  .828)  &\!\!\!& .031  & (-.778;  .871)   &\!\!\!& .010 & (-.918; .933) \\
\tau_{s_0}    &\!\!\!&-.132 &(-.242; -.021)  &\!\!\!& -.139 & (-.247;  -.034) &\!\!\!& -.128 & (-.229; -.020) \\
\hline
\vspace{-0.4cm}\\
\multicolumn{10}{l}{h=1\,500}\\
\Pr(G_i=AA)   &\!\!\!& .332 &  (.315;   .349) &\!\!\!&  .332 &  (.326;  .337) &\!\!\!& .329 &( .312;  .346) \\
\Pr(G_i=CA)    &\!\!\!& .042 &  (.035;   .077) &\!\!\!&  .027 &  (.020;  .066) &\!\!\!& .042 &( .036;  .062) \\
\Pr(G_i=NA)    &\!\!\!& .625 &  (.591;   .642) &\!\!\!&  .642 &  (.605;  .644) &\!\!\!& .628 &( .606;  .646) \\
\tau_{AA,s_0} &\!\!\!&-.183 & (-.286;  -.077) &\!\!\!& -.187 & (-.291; -.085) &\!\!\!&-.153 &(-.247; -.063) \\
\tau_{CA,s_0} &\!\!\!& .010 & (-.304;   .797) &\!\!\!&  .011 & (-.207;  .928) &\!\!\!& .000 &(-.154;  .951) \\
\tau_{s_0}   &\!\!\!&-.153 & (-.256;  -.040) &\!\!\!& -.165 & (-.266; -.057) &\!\!\!&-.130 &(-.217; -.019) \\
\hline
\end{array}
\]
}}
 \end{table}

\subsection{Posterior Predictive Model Checking} \label{sec:PPC} 
Assessing the plausibility of model assumptions is critical in model-based approaches. Model checking here is not as crucial as in other model-based approaches thanks to the randomization
 assumption, but it is still prudent to check the model fit since there
 are uncertainties in the selection of $\Uset$. We adopt Bayesian posterior predictive checks \citep[][]{Gelman96} to assess goodness-of-fit of our models in the application. Posterior predictive checks evaluate goodness-of-fit of models by measuring the discrepancy between the observed data and replicated data simulated from its posterior predictive distribution. The particular procedure adopted here is similar to that in \citet[][Section 6]{MLM13}. Specifically, we consider three discrepancy measures aim at assessing whether the model can preserve broad features of signal, noise and signal-to-noise ratio (SNR) in the drop-out status distribution for compliant-applicants, always-applicants  and the union of these two principal strata, and calculate posterior predictive $p-$values (PPPVs) to summarize discrepancies between the observed data and replicated data. Extreme (close to 0 or 1) PPPVs can be interpreted as evidence of lack-of-fit of the model in, at least some aspects of, the observed data. Further details of the procedure are relegated in \ref{websupp}.

 Table \ref{tab:PPPVs} shows the PPPVs for the model-fit to the subpopulation with bandwidth of $500$, $1\,000$ and $1\,500$ euros, respectively. The PPPVs suggest good model-fit for all bandwidths, except for a slight under-fit for always-applicants in the subpopulation with $h=500$, which is possibly due to the small sample size. We have also calculated the less conservative sampled posterior predictive $p-$values \citep{Johnson07, Gosselin11} and obtained similar conclusions. 
\begin{table}
\centering
 \caption{Bayesian $p-$values of signal, noise and SNR under different $h$ for the model used in the application to Italian university grants.} \label{tab:PPPVs}
 \begin{tabular}{llcccc}
 \hline
 h  &{Principal strata} && Signal & Noise & SNR \\
 \hline
     &\{CA\}   &&	.095	&	.630	&	.094		\\
 500 &\{AA\}   &&	.254	&	.325	&	.254		\\
     &\{AA,CA\} &&	.338	&	.273	&	.370		\\
 \hline
     &\{CA\}   &&	.411	&	.425	&	.419		\\
 1000&\{AA\}   &&	.400	&	.444	&	.473		\\
     &\{AA,CA\} && .493	&	.335	&	.518		\\
 \hline
     &\{CA\}   &&	.208	&	.444	&	.210		\\
 1500&\{AA\}   &&	.372	&	.400	&	.261		\\
     &\{AA,CA\} && .455	&	.337	&	.470		\\
 \hline
 \end{tabular}
 \end{table}

\section{Discussion} \label{sec:Conclusion}
Motivated from the evaluation of Italian university grants, we propose a probabilistic formulation of the assignment mechanism for regression discontinuity designs and develop a full Bayesian approach to draw causal inference within the framework of principal stratification. In particular, we illustrate how to utilize information on application status to gain additional insights in program evaluation. Applying the method to the data from two Italian universities, we find university grants reduce dropping out of higher education for students from low-income families and the effect size is especially pronounced for motivated students (always-applicants).

The core of the approach we propose is the local randomization assumption (Assumption~\ref{Ass_randomization}),
which is intrinsically non-testable.   Therefore, it may be worthwhile to conduct sensitivity analyses
aimed at assessing the robustness of the results with respect to violations of the local randomization assumption.
To this end, we conduct further analyses deriving the posterior distributions of the causal estimands of interest under three additional model specifications:
$(1)$ a model where we specify  the model for principal strata, $\pi_{i,g}$, and  the conditional model for potential outcomes given principal strata, $f_{i, gz}$, conditioning on neither the forcing variable nor the pre-treatment variables; $(2)$ a model where we specify  $\pi_{i,g}$ and $f_{i, gz}$  conditioning only on the forcing variable, without including the pre-treatment variables; and $(3)$ a model where we specify  $\pi_{i,g}$ and $f_{i, gz}$  conditioning only on the pre-treatment
variables, without including the forcing variable.
Under local randomization, adjusting inference for either the forcing variable, $S$, or the pre-treatment variables, $\bX$,  should not be required, therefore we expect that results are similar across different model specifications. Indeed results, shown in \ref{websupp}, are robust across different model specifications, suggesting that causal inference under the local randomization assumption is credible and fully defensible.

A fundamental distinction between our approach and the previous local-regression based RD approaches lies in the role of the forcing variable in the analysis. Specifically, previous approaches generally view the forcing variable as a pre-assignment covariate rather than a random variable as in our approach. As a consequence, the standard overlap assumption, which requires that there are both treated and control units for all values of the covariates including the forcing variable, is violated. Violation of the overlap assumption implies that the conditional independence assumption, which trivially holds in RD settings, cannot be exploited directly. Instead some kind of extrapolation is required, and in order to avoid that estimates heavily rely on extrapolation, previous analyses focus on causal effects of the treatment for units \textit{at} the threshold. Smoothness assumptions, for example, continuity of conditional regression functions of potential outcomes given the forcing variable, are usually assumed to draw inference on those causal effects. Local randomization and continuity are different assumptions, leading to different causal estimands: under continuity assumptions units with a realized value of the forcing variable around the threshold are used to draw inference on causal effects for units \textit{at} the threshold, whereas under local randomization, inference is drawn on causal effects for units \textit{around} the threshold.

In the evaluation of Italian university grants, other than dropout, student's academic performance (measured by total credits taken or passing rate of exams) is also of great interest in policy. As illustrated by \cite{MLM13} and \cite{Mercatanti14}, jointly modeling two outcomes, dropout and academic performance in this case, would be worthwhile for both practical and inferential purposes, and it is at the top of our research agenda.

After the first year, the Italian university grant assignment rule combines sequential and RD designs \citep{Cellini10}: grants are allocated both on the basis of students family economic indicator and on the ground of their academic performance (exam scores above a certain threshold). Such complex assignment mechanisms pose challenges to causal inference, requiring new structures and assumptions; meanwhile, they also present great opportunities for extending the existing framework to more general RD settings. One specific direction of our future research is to develop methods that combine Bayesian tools for RDs and dynamic treatment regimes \citep{Murphy03,Zajonc12} in the presence of multiple forcing variables \citep{ImbensZajonc11}.

\section*{Acknowledgements}
The authors are grateful to the associate editor and two reviewers for constructive comments, to Quanli Wang for computing support, to Guido Imbens, Michael Hudgens and Sid Chib for helpful discussions.

 \begin{supplement}[id=websupp]
   \sname{Web Supplementary Material}
   \stitle{Details of Caculation and Sensitivity Analysis}
   \slink[url]{http://lib.stat.cmu.edu/aoas/???/???}
   \sdescription{We describe in detail the Bayesian approach we used to select the subpopulations,
      the Markov Chain Monte Carlo (MCMC) methods used to simulate the posterior distributions of the parameters of the models, the posterior predictive checks, and the sensitivity analysis regarding local randomization described in Section \ref{sec:Conclusion}.}
 \end{supplement}

\bibliographystyle{imsart-nameyear}
\bibliography{RDD_AOAS}
\end{document}